\documentclass{article}

\usepackage{arxiv}

\usepackage{natbib}

\usepackage{amsmath, amsfonts, amssymb, amsthm, dsfont, mathtools, bm}

\usepackage[utf8]{inputenc} % allow utf-8 input
\usepackage[T1]{fontenc}    % use 8-bit T1 fonts
\usepackage[colorlinks=true, linkcolor=red, urlcolor=blue, citecolor=blue, psdextra, pdfencoding=auto]{hyperref}
\usepackage{url}            % simple URL typesetting
\usepackage{booktabs}       % professional-quality tables
\usepackage{nicefrac}       % compact symbols for 1/2, etc.
\usepackage{microtype}      % microtypography
\usepackage{lipsum}
\usepackage{graphicx}
\graphicspath{ {./images/} }

\newtheoremstyle{mytheoremstyle} % name
    {0.3cm}                      % Space above
    {0cm}                        % Space below
    {\itshape}                   % Body font
    {}                           % Indent amount
    {\scshape}                   % Theorem head font
    {: }                          % Punctuation after theorem head
    {0em}                       % Space after theorem head
    {}  % Theorem head spec (can be left empty, meaning normal)

\theoremstyle{mytheoremstyle}

\newtheorem*{Lemma*}{Lemma}

\newtheorem{Proposition}{Proposition}

\renewenvironment{proof}{{\noindent \sc Proof:}}{\qed}

\newtheoremstyle{myExampleRemarkstyle} % name
    {0.3cm}                    % Space above
    {0cm}                           % Space below
    {\itshape}                   % Body font
    {}                           % Indent amount
    {\scshape}                   % Theorem head font
    {: }                          % Punctuation after theorem head
    {0em}                       % Space after theorem head
    {}  % Theorem head spec (can be left empty, meaning normal)

\theoremstyle{myExampleRemarkstyle}
 
\newtheorem{Remark}{Remark}
\newtheorem{Assumption}{Assumption}

\DeclareMathOperator*{\argmax}{argmax}
\DeclareMathOperator*{\argmin}{argmin}

\DeclareMathOperator*{\var}{var}
\def\bgam{\bm{\gamma}}
\DeclarePairedDelimiter\ceil{\lceil}{\rceil}

\title{Prevalence Estimation from Random Samples and Census Data with Participation Bias}

\author{
    St\'ephane Guerrier\\
    Geneva School of Economics and Management \&\\ 
    Faculty of Science\\
    University of Geneva, Switzerland,\\
    \texttt{stephane.guerrier@unige.ch} \\
   \And
     Christoph Kuzmics\\
    Department of Economics,\\
    University of Graz, Austria\\
    \texttt{christoph.kuzmics@uni-graz.at} \\
  \And
    Maria-Pia Victoria-Feser \\
    Geneva School of Economics and Management\\ 
    University of Geneva, Switzerland,\\
    \texttt{maria-pia.victoriafeser@unige.ch} \\
}

\begin{document}
\maketitle
\begin{abstract}
Countries officially record the number of COVID-19 cases based on medical tests of a subset of the population with unknown participation bias. For prevalence estimation, the official information is typically discarded and, instead, random survey samples are taken. An exception is the surveys recorded by the Statistics Austria Federal Institute, were the sample contains information about the number of positive COVID-19 tests in the sample as well as the participants with a positive COVID-19 test measured through the official procedure in the population during the same period. We derive (maximum likelihood and method of moment) prevalence estimators, with possible measurement errors, based on a survey sample, that additionally utilize the official information. We show that they are substantially more accurate than the simple survey sample proportion of positive cases. Put differently, using the proposed estimators, the same level of precision can be obtained with substantially smaller survey sample sizes. Moreover, the proposed estimators are less sensitive to measurement errors due to the sensitivity and specificity of the medical testing procedure. The proposed estimators and associated confidence intervals are implemented in the companion open source R package \texttt{cape}.
\end{abstract}

% keywords can be removed

\section{Introduction}
\label{sec1}

In the ongoing COVID-19 pandemic, governments face a trade-off between reducing the wealth or the health of citizens when choosing the degree of economic slowdown in their policy measures. The key to assess this trade-off is an understanding of the number or proportion of cases in the population and their evolution. Acquiring this understanding, in turn, depends on reliable estimates of the number of cases (at different points in time). 

The officially recorded number of positive cases can probably only be seen as a lower bound of the actual number of cases. The selection of participants to be medically tested is typically not complete and, importantly, also not random, but instead suffers from an unknown participation bias. The whole official procedure can, in fact, be understood as a complete census with (a possibly large) participation bias. It is typically unclear how many undetected positive cases there are in the population. Acknowledging this problem, for the case of COVID-19, some studies have proposed estimates for the prevalence among asymptomatic patients (see e.g. \citealp{NISHIURA2020}, \citealp{Mizumoto2020}), or have attempted to infer from the prevalence obtained through the official procedure to the population one \citep[see e.g.][]{MANSKI2020}.

In this paper, we instead propose to combine the information available in the data obtained through the official procedure that, as argued, suffers from participation bias, with data collected using a random  sample of participants from the population all of which are medically tested. From this random sample, an unbiased estimator of the population proportion of positive cases, ignoring the information available from the official procedure, is then simply the  proportion of positive cases in the sample; see e.g. \cite{Bendavid2020,X2MIHW_2020,Stringhini2020} for the analysis of COVID-19 prevalence. 
More precisely, we demonstrate that the information gathered through the official procedure, while not  useful in its own, can be used to improve the accuracy of the best estimators derived from random samples. All what is needed, is to also record, for each participant in the random sample, whether they are already part of the official statistics, i.e., whether they have been already declared positive through the official procedure. Appropriate estimators can then be derived whose key input is the number of new cases found in the sample.

We show that these estimators are substantially more accurate than the standard proportion of cases in the random sample. Or put differently, appropriately utilizing the information obtained through the official procedure, means that the sample sizes for the survey can be substantially smaller and yet achieve the same statistical accuracy, thus, substantially reducing the costs and/or time for data acquisition. Alternatively, from the same survey sample, finer analysis at sub-population levels (e.g. regions) can reasonably be done even if the number of participants in these levels is rather small.

We also provide several standard approaches to building confidence interval bounds for the proportion of positive cases, and compare, in a simulation study, their (finite sample) coverage properties. We also take into account possible misclassification errors of the (medical) testing devices used to collect the data %, such as Rapid Diagnostic Test (RDT) or serological tests such as Enzyme-Linked ImmunoSorbent Assai (ELISA) 
see e.g., \cite{KoWeGr:20} and \cite{surkova2020false}. The associated misclassification errors are actually induced by their sensitivity, i.e., the complement to the False Positive (FP) rate, and by their specificity, i.e., the complement to the False Negative (FN) rate, and adjusting for these errors avoids biased estimates \citep[see e.g.][and the references therein]{Digg:11,Lewi:12}. Using a sensitivity analysis with the Austrian survey data, we actually find that the proposed estimators are much less influenced by the value of the FN rate, than the survey sample proportion, allowing, in practice, to limit the impact of the choice for the medical test specificity when estimating the proportion of positive cases. 

Such misclassification adjustments are also necessary with binary outcomes in logistic regression; see e.g. \cite{NiDaKaTaLe:19} and  \cite{MEYER2017}, and the references therein. In this paper, we consider the case of estimating the proportion of positive cases, but the framework could easily be extended to the case of logistic regression. Moreover, while the data from the November 2020 survey collected by \cite{StatAU:20} is suitable for prevalence estimation, i.e. the population proportion of Austrians infected by the COVID-19 in November 2020, the same approach can be used to estimate other proportions such as the incidence of the COVID-19 \citep[see e.g.][]{EpidBook}. For the sensitivity and specificity, we use cutoff values, hence without the need to specify a (prior) distribution for these quantities \citep[see e.g.][and the references therein]{McDoHo:18,Bouman2020}. %However, our estimation framework can, in principle, also be extended to these settings. 
Finally, the data from the Austrian survey \citep{StatAU:20} is performed using the \texttt{cape} R package which includes the new methods developed in this paper (see Section \ref{sec:cape} for more details).

The paper is organised as follows. We first present the formal setup in Section~\ref{sec:setup}. In Section~\ref{sec:general} we derive associated estimators and inference procedures, also treating the case of possible (partially) missing information. In Section~\ref{sec:simu} we present a simulation study that confirms and quantifies the theoretical results we develop in the previous sections. In Section~\ref{sec:case-study} we apply the methodology to the case of the COVID-19 prevalence estimation and associated confidence bounds in Austria.

\section{The Model}
\label{sec:setup}

Consider taking a (random) survey sample of $n$ participants in some population in order to estimate the population proportion $\pi$ of, for example, a given infectious disease. Our framework also supposes that prior to the collection of the survey sample, a known proportion of individuals in the population have been declared positive through an official procedure based on an incomplete census or a census with participation bias. The official procedure has two steps. First, participants are selected based on some unknown criteria. Second, selected participants are medically tested for the disease.  

For each participant $i=1,\ldots,n$ in the survey sample, there are three random variables of interest.
\begin{equation}
    \begin{aligned}
         X_i&\vcentcolon=
         \left\{
	\begin{array}{ll}
		1  & \quad \mbox{if participant $i$ is positive,} \\
		0  & \quad \mbox{otherwise;}
	\end{array}
\right.\\
         Y_i&\vcentcolon=
         \left\{
	\begin{array}{ll}
		1  & \quad \mbox{if participant $i$ is tested positive in the survey sample,} \\
		0  & \quad \mbox{otherwise;}
	\end{array}
\right.\\
         Z_i&\vcentcolon= \left\{
	\begin{array}{ll}
		1  & \quad \mbox{if participant $i$ was declared positive with the official procedure,} \\
		0  & \quad \mbox{otherwise.}
	\end{array}
\right. 
    \end{aligned}
    \label{eqn:bernoulli-rv}
\end{equation}
We assume that, for each participant $i=1,\ldots,n$ in the survey sample, we observe $Y_i$ and $Z_i$, but not $X_i$. The objective is to provide an estimator for the unknown population proportion
\begin{equation*}
        \pi \vcentcolon= \mathbb{P}\left(X_i=1\right). 
\end{equation*}
We allow for the possibility that the outcome of (medical) tests can be subject to misclassification error. Let
\begin{equation*}
    \begin{aligned}
          \pi_0 &\vcentcolon= \mathbb{P}(Z_i=1), && \\
    \alpha_0&\vcentcolon= \mathbb{P}(Z_i=1\vert X_i=0), & \quad \quad \alpha&\vcentcolon= \mathbb{P}(Y_i=1\vert X_i=0), \\
    \beta_0&\vcentcolon= \mathbb{P}(Z_i=0\vert X_i=1), & \quad \quad \beta&\vcentcolon= \mathbb{P}(Y_i=0\vert X_i=1).
    \end{aligned}
\end{equation*} \medskip

The probabilities $\alpha$ and $\beta$, are the (assumed known) FP rates ($\alpha = 1-\mbox{specificity}$) and FN rates ($\beta = 1-\mbox{sensitivity}$) of the particular medical test employed in the survey. The probabilities $\alpha_0$ and $\beta_0$, are respectively the (assumed known) FP and FN rates of the official procedure.  

\begin{Remark} \label{rem:alpha0}
{\it Note that $\alpha_0$ is not the FP rate of the medical test administered in the official procedure. It is the probability that a participant has been incorrectly declared positive through the official procedure and, therefore, the product of two probabilities: the probability that a negative individual is selected to be tested in the official procedure multiplied with the probability that the medical test is positive conditional on this individual being (selected and) negative. In many applications $\alpha_0$ will, therefore, be, sometimes substantially, smaller than the FP rate of the medical test.} 
\end{Remark}

The FN rate $\beta_0$ of the official procedure is not known (otherwise we would know the population proportion $\pi$) and depends on $\pi_0$, $\pi$ and $\alpha_0$ as follows: 
\begin{eqnarray*}
\pi_0 = \mathbb{P}(Z_i=1) & = & \mathbb{P}(X_i=0) \mathbb{P}(Z_i=1 \vert X_i=0) + \mathbb{P}(X_i=1) \mathbb{P}(Z_i=1 \vert X_i=1) \\
& = & (1-\pi) \alpha_0 + \pi (1-\beta_0), 
\end{eqnarray*}
Thus, 
\[
\beta_0 = 1- \frac{\pi_0-\alpha_0(1-\pi)}{\pi}.
\]
It is useful to make three small assumptions. 
\begin{Assumption}
\label{ass:alpha-beta}
$\alpha+\beta<1$. 
\end{Assumption}
\begin{Assumption}
\label{ass:alpha0}
$\alpha_0 + \beta_0 < 1$.
\end{Assumption}
\begin{Assumption}
\label{ass:sample-drawing}
The survey sample is collected completely at random, without replacement. Its size $n$ is small compared to the population size.
\end{Assumption}
With Assumption \ref{ass:alpha-beta}, we rule out the uninteresting case $\alpha+\beta = 1$. Indeed, if $\alpha+\beta=1$, $Y_i$ would be completely uninformative about the random variable of interest $X_i$, as $\mathbb{P}(X_i=1|Y_i=1) = \mathbb{P}(X_i=1|Y_i=0) = \pi$. Otherwise Assumption \ref{ass:alpha-beta} is without loss of generality in the following sense. If $\alpha+\beta>1$, we could just use $Y_i'=1-Y_i$ instead of $Y_i$, which would have FP and FN rates of $\alpha'=1-\alpha$ and $\beta'=1-\beta$, with $\alpha'+\beta'<1$.

Assumption \ref{ass:alpha0} is similarly without loss of generality. It implies that $\alpha_0 \le \pi_0$. To see this suppose that $\alpha_0 > \pi_0 = (1-\pi) \alpha_0 + \pi (1-\beta_0)$. This is equivalent to $0 > -\pi \alpha_0 + \pi (1-\beta_0)$, which in turn, is equivalent to $0 > 1 - \alpha_0 - \beta_0$, a contradiction.  

Assumption~\ref{ass:sample-drawing} specifies the type of sampling method assumed in this paper. Extensions to weighted sampling methods, with non random weights, would require a relatively straightforward adjustment of the proposed estimators, that we omit for clarity of exposition. Moreover, assuming that the sample size is relatively small compared to the population size, allows one to consider distributional properties of the variables that can be easily defined, in that binomial distributions can be used to approximate hypergeometric distributions. 

\begin{Remark}\label{rem:lowbound}
The unknown population proportion of positive cases $\pi$ is bounded from below by $\underline{\pi} \vcentcolon= \frac{\pi_0-\alpha_0}{1-\alpha_0}$. To see this, recall that the equality $\pi_0 = (1-\pi) \alpha_0 + \pi (1-\beta_0)$ must hold (with both $\pi$ and $\beta_0$ unknown parameters). %This equation can be satisfied for $\pi=1$, which would imply that $\pi_0=1-\beta_0$. 
The lowest admissible value for $\pi$ is achieved when $\beta_0=0$, in which case we get the lower bound $\frac{\pi_0-\alpha_0}{1-\alpha_0}$. If $\alpha_0=0$ then $\underline{\pi}=\pi_0$. Note that, given the assumptions, $0 \le \frac{\pi_0-\alpha_0}{1-\alpha_0} \le 1$. 
\end{Remark}

From these variables we construct the following random variables that will be used to formulate the models:
\begin{alignat}{4}
    &R_{11}&&\vcentcolon=\sum_{i=1}^n Y_i Z_i, \quad\quad\quad 
    &&R_{10}&&\vcentcolon=\sum_{i=1}^n(1-Y_i)Z_i, \label{eqn:R-R4} \\
    &R_{01}&&\vcentcolon= \sum_{i=1}^nY_i(1-Z_i), \quad\quad \quad 
    &&R_{00}&&\vcentcolon= \sum_{i=1}^n(1-Y_i)(1-Z_i)=n-R_{11}-R_{10}+R_{01}.
    \nonumber
\end{alignat}
In words, $R_{11}$ is the number of participants in the survey sample that are tested positive  and have also been declared positive through the official procedure; $R_{10}$ is the number of participants in the survey sample that are tested negative but have been declared positive through the official procedure; $R_{01}$ is the number of participants in the survey sample that are tested positive but have been declared negative through the official procedure; $R_{00}$ is the number of participants in the survey sample that are tested negative and have been declared negative through the official procedure. 
We also make use of $R_{\ast 1} = \sum_{i=1}^nY_i = R_{11} + R_{01}$, the number of participants that are tested positive in the survey sample. 

The success probabilities (see Supplementary Material \ref{app:succprob} for their derivation), denoted by $\tau_{ij}(\pi)$ associated to each $R_{ij}$, $i,j \in \{0,1\}$ in \eqref{eqn:R-R4} are given by 
\begin{eqnarray}
    \begin{aligned}
    \tau_{11}(\pi) &\vcentcolon=  \mathbb{P}(Z_i=1, Y_i=1) =\pi\Delta\alpha_0+(\pi_0-\alpha_0)(1-\beta)+\alpha\alpha_0,   \\
    \tau_{10}(\pi) &\vcentcolon=  \mathbb{P}(Z_i=1,Y_i=0) = -\pi\Delta\alpha_0+(\pi_0-\alpha_0)\beta+(1-\alpha)\alpha_0,  \\
    \tau_{01}(\pi) &\vcentcolon=  \mathbb{P}(Z_i=0,Y_i=1) =  \pi\Delta(1-\alpha_0)-(\pi_0-\alpha_0)(1-\beta)+\alpha(1-\alpha_0), \\
    \tau_{00}(\pi) &\vcentcolon=  \mathbb{P}(Z_i=0,Y_i=0) = -\pi\Delta(1-\alpha_0)-(\pi_0-\alpha_0)\beta+(1-\alpha)(1-\alpha_0),
    \end{aligned}
    \label{eqn:tau-pi}
\end{eqnarray} 
where $\Delta \vcentcolon=1-(\alpha+\beta)$. Without misclassification error, we would have $\tau_{11}(\pi)=\pi_0$, $\tau_{10}(\pi)=0$, $\tau_{01}(\pi)=\pi-\pi_0$, $\tau_{00}(\pi)=1-\pi$. Moreover, it is easy to verify that given our Assumptions, we have that the $\tau$'s are non-negative and sum up to $1$.

\section{Estimation and Inference} \label{sec:general}

In this section we derive Maximum Likelihood Estimators (MLE), a marginal MLE when some data is missing, and some Generalized Method of Moment (GMM) estimators. We also provide (exact) fiducial confidence intervals when possible, such as for a Method of Moment Estimator (MME) estimator under the assumption that the FP rates are zero. We also provide confidence intervals based on the estimators' asymptotic distribution. We compare the accuracy of the proposed estimators (that utilize the information from the official procedure) with the survey MLE that is the sample proportion of positive cases in the survey sample (that ignores the information from the official procedure). 

\subsection{Estimators}
\label{sec:estim-ME}

%\subsubsection{Survey sample proportion estimator}
\subsubsection{Survey MLE}

The benchmark estimator which is based only on $R_{\ast 1}(=R_{11}+R_{01})$, the number of positive cases in the survey sample, is given by %
\begin{equation}
    \bar{\pi}=\frac{R_{\ast 1}/n-\alpha}{\Delta},
    \label{eqn:naive-ME}
\end{equation}
which reduces to $\bar{\pi}=R_{\ast 1}/n$, when $\alpha=\beta=0$. It is actually the MLE of $\pi$ based only on the survey sample.

Its variance is given by 
\begin{equation} \label{eqn:var_naive}
     \var(\bar{\pi}) = \frac{(\tau_{11}(\pi)+\tau_{01}(\pi))(1-\tau_{11}(\pi)-\tau_{01}(\pi))}{n\Delta^2} = \frac{(\pi\Delta+\alpha)(1-\pi\Delta-\alpha)}{n\Delta^2}.
\end{equation}

\subsubsection{Conditional MLE}

Under Assumption \ref{ass:sample-drawing}, the likelihood function for $\pi$ can be obtained from the multinomial distribution with categories provided by $R_{11},R_{10},R_{01},R_{00}$ and their associated success probabilities $\tau_{11}(\pi),\tau_{10}(\pi),\tau_{01}(\pi),\tau_{00}(\pi)$. The log-likelihood function is, therefore, given by
\begin{equation}
        \ell(\pi\vert R_{11},R_{10},R_{01},R_{00}) =  C + \sum_{i=1}^1 \sum_{j=0}^1 R_{ij} \ln(\tau_{ij}(\pi)),
        \label{eqn:loglik-MCE}
\end{equation}
where $C$ is a quantity independent of $\pi$. 

The conditional MLE, i.e., the one based on the log-likelihood given in \eqref{eqn:loglik-MCE}, which is hence conditional on the information provided by the official procedure, is defined by
\begin{equation}
       \hat{\pi} \vcentcolon= \argmax_{\pi \in [\underline{\pi},1]} \;  \ell(\pi\vert R_{11},R_{10},R_{01},R_{00}),
       \label{eqn:MLE-ME}
\end{equation}
with $\underline{\pi}$ given in Remark \ref{rem:lowbound}. The conditional MLE $\hat{\pi}$, generally, has no closed-form solution but can be computed numerically. However, in the case when $\alpha_0 = 0$, we obtain a closed-form solution given by
\begin{equation}
    \hat{\pi} = \frac{\pi_0 R_{00} + R_{01}}{\Delta \left(R_{01}+R_{00}\right)} - \frac{\pi_0 \beta}{\Delta} - \frac{\alpha}{\Delta}.
    \label{eq:cmle-closedform}
\end{equation}
When $\alpha_0=\alpha=\beta=0$, this further reduces to 
\begin{equation}
   \hat{\pi} = \pi_0 \frac{n - R_{\ast 1}}{n - R_{11}} + \frac{R_{01}}{ \left(n - R_{11}\right)}.
   \label{eqn:MLE}
\end{equation}

\begin{Remark} \label{rem:interior}
{\it The closed form expression in \eqref{eq:cmle-closedform} is the conditional MLE only if the estimate is within the interval $[\underline{\pi},1]$. There are, however, possible (but unlikely in practice) combinations of parameter values and sample realisations for which the likelihood function is maximized at the boundaries, i.e. either at $\underline{\pi}$ or at $1$. In the case of no misclassification errors ($\alpha_0=\alpha=\beta=0$) the estimate given in \ref{eqn:MLE} is automatically within $[\underline{\pi},1]$.}
\end{Remark}

\begin{Remark} \label{rem:conditional}
{\it When $\alpha_0=\alpha=\beta=0$, in \eqref{eqn:bernoulli-rv}, $Y_i=X_i$ and $Z_i \le X_i$, so that, in \eqref{eqn:R-R4}, $R_{10}=0$, $R_{01}=\sum_{i=1}^n Z_i$, and $R_{*1}=\sum_{i=1}^n X_i$ with $R_{01} \leq R_{*1}$. Under Assumption \ref{ass:sample-drawing}, we have that $R_{*1}\sim\mathcal{B}(n,\pi)$ and, conditionally on $R_{*1}$, we obtain the conditional model $R_{01}\vert R_{*1}\sim\mathcal{B}(R_{*1},\frac{\pi_0}{\pi})$. The associated (conditional) likelihood function is, therefore, given by
\begin{equation*}
    \mathcal{L}(\pi\vert R_{01},R_{*1}) = {n \choose R_{*1}} (\pi)^{R_{*1}}\left(1- \pi\right)^{(n-R_{*1})} {r \choose R_{01}} \left(\frac{\pi_0}{\pi}\right)^{R_{01}} \left(1-\frac{\pi_0}{\pi}\right)^{(R_{*1}-R_{01})},
\end{equation*}
with associated conditional MLE given in \eqref{eqn:MLE}.
}
\end{Remark}

In Proposition \ref{prop:mle} below, we show the consistency and asymptotic normality of the conditional MLE defined in \eqref{eqn:MLE-ME}. 

\begin{Proposition}
\label{prop:mle}
The conditional MLE $\hat{\pi}$ defined in \eqref{eqn:MLE-ME} is  consistent for $\pi$. Moreover, if $\pi \in (\underline{\pi}, 1)$, we have
\begin{equation*}
    \sqrt{n} \left(\hat{\pi} - \pi\right) \xrightarrow[n \to \infty]{\mathcal{D}} \mathcal{N} \left(0, \frac{1}{I(\pi)}\right),
\end{equation*}
where 
\begin{equation*}
    I(\pi) = \left\{ \begin{array}{cc} \displaystyle \frac{\left(\frac{d \tau_{11}(\pi)}{d \pi}\right)^2}{\tau_{11}(\pi)} +  \sum_{j=0}^1 \frac{\left(\frac{d \tau_{0j}(\pi)}{d \pi}\right)^2}{\tau_{0j}(\pi)} & \mbox{ if } \alpha_0 = \beta = 0 \\
    \displaystyle \sum_{i=0}^1 \sum_{j=0}^1 \frac{\left(\frac{d \tau_{ij}(\pi)}{d \pi}\right)^2}{\tau_{ij}(\pi)} & \mbox{ otherwise. }
    \end{array}\right.
\end{equation*}
\end{Proposition} \bigskip

The proof of Proposition \ref{prop:mle} is provided in Supplementary Material \ref{app:proof-prop:mle}. 

\subsubsection{GMM Estimators}

Alternatively, we can consider an estimator from the class GMM estimators \citep{Hans:82} based on the random variable $\mathbf{R} \vcentcolon=[R_{11}/n, R_{10}/n, R_{01}/n, R_{00}/n]$ with expectation $\mathbb{E}[\mathbf{R}]:=\bm{\tau}(\pi)=[\tau_{11}(\pi), \tau_{10}(\pi), \tau_{01}(\pi), \tau_{00}(\pi)]$. A GMM estimator $\ddot{\pi}$ is given by
\begin{equation*}
    \ddot{\pi} \vcentcolon=\argmin_{\pi\in [\underline{\pi},1]}Q(\pi\vert \mathbf{R}),
\end{equation*}
with
\begin{equation*}
    Q(\pi\vert \mathbf{R})\vcentcolon=\left(\mathbf{R}-\bm{\tau}(\pi)\right)^T\bm{\Omega}\left(\mathbf{R} - \bm{\tau}(\pi)\right),
\end{equation*}
where $\bm{\Omega}$ is a fixed 4 by 4 positive definite matrix with entries $\omega_{ij}$, $i,j=1,...,4$. Since 
$\bm{\tau}(\pi)$ is a linear combination of $\pi$, we can write $\bm{\tau}(\pi):=\mathbf{a}\pi+\mathbf{b}$, with 
$\mathbf{a}=[a_l]_{l=1,\ldots,4}$, $\mathbf{b}=[b_l]_{l=1,\ldots,4}$ two  vectors derived from \eqref{eqn:tau-pi}. Then, assuming an interior solution exists (a remark similar to Remark \ref{rem:interior} applies), $\ddot{\pi}$ is the root of
\begin{equation*}
    {\frac{d}{d\pi}} \, Q(\pi\vert \mathbf{R})=-2\left(\mathbf{R}-\bm{\tau}(\pi)\right)^T\bm{\Omega}\mathbf{a}. 
\end{equation*}
Therefore, we obtain 
\begin{equation}
    \ddot{\pi}=\frac{(\mathbf{R}-\mathbf{b})^T\bm{\Omega}\mathbf{a}}{\mathbf{a}^T\bm{\Omega}\mathbf{a}},
    \label{eqn:GMM}
\end{equation}
and it follows that $\mathbb{E}[\ddot{\pi}]=\pi$. For a general matrix $\bm{\Omega}$, $\ddot{\pi}$ is a linear combination of the elements of $\mathbf{R}$, and it would be useful to choose $\bm{\Omega}$ such that the distribution of $\ddot{\pi}$ is known (for all $n$), for the construction of exact confidence bounds. One such case is obtained when $\omega_{ij}=1$ for $i=j=3$ and $0$ otherwise, i.e. the GMM is reduced to a MME based on $R_{01}$ (with expectation $\tau_{01}(\pi)$), which, again assuming an interior solution exists, is given by $\tilde{\pi} \in [\underline{\pi},1]$ that solves
\begin{equation*}
     \tau_{01}(\tilde{\pi}) = \frac{R_{01}}{n}.
\end{equation*}
This yields
\begin{equation}
    \tilde{\pi} = \frac{1}{\Delta(1-\alpha_0)} \left(\frac{R_{01}}{n} + \pi_0 - \beta\pi_0 - \alpha_0 \Delta - \alpha\right).
    \label{eqn:MME-ME}
\end{equation}
When $\alpha_0=\alpha=\beta=0$, this reduces to 
\begin{equation}
   \tilde{\pi} = \pi_0 + \frac{R_{01}}{n}.
    \label{eqn:MME}
\end{equation}

\begin{Remark}
{\it Interestingly, in the case of no misclassification errors ($\alpha_0=\alpha=\beta=0$), $\tilde{\pi}$ can also be seen as an approximation to the MLE (in \ref{eqn:MLE}) for small values of $\pi_0$ and $\pi$, i.e., by simplifying $(n-R_{\ast 1})/(n-R_{11})\approx 1$ and $\pi_0(n - R_{11}) \approx \pi_0 n$.}
\end{Remark}

Moreover, we have $\mathbb{E}[\tilde{\pi} ] = \pi$, i.e., the moment estimator is unbiased, and the variance is easily determined to be
\begin{equation} \label{eqn:var_MME}
\var\left(\tilde{\pi}\right) =  \frac{1}{\Delta^2(1-\alpha_0)^2} \var \left(\frac{R_{01}}{n}\right) =  \frac{\tau_{01}(\pi) (1 - \tau_{01}(\pi))}{n\Delta^2(1-\alpha_0)^2} .
\end{equation}
The possible advantage of the MME $\tilde\pi$ in \eqref{eqn:MME-ME} is that is has a known finite sample distribution, based on $R_{01}\sim\mathcal{B}(n,\tau_{01}(\pi))$, so that exact confidence bounds can be computed using, for example, the Clopper-Pearson method, see below. Actually, using \eqref{eqn:GMM} and setting  $\omega_{ij}=1$ for $i=j=l$ and $0$ otherwise, $l=1,\ldots,4$, we can obtain all the MME corresponding to the different variables in $\mathbf{R}$, as
\begin{equation}
    \tilde{\pi}^{(l)}=\frac{R_l/n-b_l}{a_l},
    \label{eqn:pi-l}
\end{equation}
with $\mathbb{E}[\tilde{\pi}^{(l)}]=\pi$ for all $l=1,\ldots,4$, and also known finite sample distribution. In Supplementary Material \ref{app:gmm} we propose an alternative and more efficient moment estimator based on a (variance minimizing) linear combination of the $\tilde{\pi}^{(l)}$, but unfortunately without known finite sample distribution. However, when $\alpha_0$ tends to zero (recall Remark \ref{rem:alpha0} for the interpretation of $\alpha_0$), this minimum variance GMM estimator is in fact the MME in \eqref{eqn:MME-ME}.

\subsubsection{Missing information}

%In some cases it might be easier (less costly) to ask only those participant that are positive in the sample ($Y_i=1$) whether they had already been declared positive in the official procedure ($Z_i=1$), especially if one needs to collect this information via a follow-up procedure. Therefore, we also propose in Section \ref{sec:estim-ME} a consistent estimator that does not require the knowledge of $R_{10}$ and $R_{00}$, which appears to have a similar efficiency level as when $R_{10}$ and $R_{00}$ are available, at least for reasonably small misclassification errors.

In some cases it might be that the information in $R_{10}$ (and $R_{00}$) in \eqref{eqn:R-R4} is not easily available, for example, when additional data is collected using follow-up procedures. In that case, one can proceed with the marginalization of the likelihood function in \eqref{eqn:loglik-MCE} on the unknown quantities, leading to 
\begin{equation*}
    \begin{aligned}
        \ell^\ast(\pi\vert R_{11},R_{01}) = &\,  C + R_{11} \ln(\tau_{11}(\pi)) + R_{01} \ln(\tau_{01}(\pi)) + \\
        & + \mathbb{E} \left[R_{10}\right] \ln(\tau_{10}(\pi)) + \left(n - R - \mathbb{E} \left[R_{10}\right] \right) \ln(\tau_{00}(\pi)) \\
        = &\, C + R_{11} \ln(\tau_{11}(\pi)) + R_{01} \ln(\tau_{01}(\pi)) + \\
        & + n \tau_{10}(\pi) \ln(\tau_{10}(\pi)) + \left(n - R -  n\tau_{10}(\pi)\right)  \ln(\tau_{00}(\pi)),
    \end{aligned}
\end{equation*}
where $C$ is a quantity independent of $\pi$. The marginal MLE is given by
\begin{equation}
       \check{\pi} \vcentcolon= \argmax_{\pi \in [\underline{\pi}, 1]} \;  \ell^\ast(\pi\vert R_{11},R_{01}),
       \label{eqn:MMLE-ME}
\end{equation}
and, generally, has no closed form. It can however be easily computed using a numerical optimisation method.

As for the conditional MLE, we show the consistency and asymptotic normality of the marginal MLE in \eqref{eqn:MMLE-ME} in Proposition \ref{prop:mle-marg} below. The proof is omitted as it follows closely the one of Proposition \ref{prop:mle}. Also, the exact expression of the asymptotic variance denoted by $I^*(\pi)^{-1}$, is not explicitly provided here but implemented in the \textrm{cape} R package (see Section \ref{sec:cape}).
\begin{Proposition}
\label{prop:mle-marg}
The marginal MLE $\check{\pi}$ in \eqref{eqn:MMLE-ME} is consistent for $\pi$. Moreover, if $\pi \in (\underline{\pi}, 1)$, we have
\begin{equation*}
    \sqrt{n} \left(\check{\pi} - \pi\right) \xrightarrow[n \to \infty]{\mathcal{D}} \mathcal{N} \left(0, \frac{1}{I^{\ast}(\pi)}\right).
\end{equation*}
\end{Proposition}

\subsection{Efficiency}
\label{sec:efficiency}

In this section, we compare the variance of the various estimators to assess their efficiency relative to the Cramer-Rao lower bound variance (that the conditional MLE achieves asymptotically) of all unbiased estimators. 

The closed form expressions for the variance are given in \eqref{eqn:var_naive} for the survey MLE and in \eqref{eqn:var_MME} for the MME. No closed form expressions of the finite sample variance of the conditional MLE and the marginal MLE are easily obtained, not even for the case of no misclassification errors.

The Cramer-Rao lower bound, which is also the asymptotic variance of the conditional MLE, is given by the reciprocal of the Fisher information, that is
\begin{equation}
I(\pi )^{-1} = -\mathbb{E}\left[\frac{\partial^2}{\partial \pi^2}\ell(\pi\vert \mathbf{R})\right]^{-1}.
\label{eqn:MLE-asvar}
\end{equation}
One can provide a lengthy closed form expression for $I(\pi)^{-1}$, see Proposition \ref{prop:mle}. In practice, based on simulations (not presented here), the sample variance appears indistinguishable from the asymptotic variance, from sample sizes of $n \geq500$. 

In Section \ref{sec:simu}, we perform a simulation study, with parameter values loosely inspired by what one might expect for estimating the COVID-19 prevalence using PCR tests, to empirically assess the efficiency of the various estimators by considering the ratio of the Cramer Rao lower bound and the variance of the estimator. In this section we formally compute efficiency ratio, in the case of no misclassification errors, in order to highlight the increased precision that we get by considering the information from the official procedure. To do so, let $\alpha_0=\alpha=\beta=0$ and consider the ratio of the variance of $\bar{\pi}$ (in \ref{eqn:var_naive}) relative to $\tilde{\pi}$ (in \ref{eqn:var_MME}): 
\begin{equation}
    \frac{\var\left(\bar{\pi}\right)}{\var\left(\tilde{\pi}\right)}=\frac{\pi(1-\pi)}{(\pi - \pi_0)(1 + \pi_0 - \pi)} = \frac{\pi(1-\pi)}{\pi(1-\pi) - \pi_0 (1 + \pi_0 - 2\pi)}.
    \label{eqn:eff-barpi-tildepi}
\end{equation}
Therefore, when $2\pi > 1 + \pi_0$ we have $\var(\bar{\pi}) < \var(\tilde{\pi})$, while when $2\pi < 1 + \pi_0$ we have $\var(\bar{\pi}) > \var(\tilde{\pi})$. A sufficient condition for the variance of the MME to be lower than the variance of the survey MLE is, therefore, that the true population proportion $\pi$ is below one half. 

On the other hand, the  efficiency of the survey MLE relative to the (asymptotic) conditional MLE, in this case, is given by
\begin{equation*}
    e(\bar{\pi})= \frac{I(\pi )^{-1}}{\var(\bar{\pi})}= \frac{\pi - \pi_0}{\pi (1 - \pi_0)} < 1,
\end{equation*}
since $\pi_0 \leq \pi < 1$. 

Moreover, since the variance of the conditional MLE is also the Cramer-Rao lower bound for the variance of any unbiased estimator of $\pi$, the MME, being unbiased, must have a higher variance. Indeed, the  relative  efficiency of $\tilde{\pi}$ versus the conditional MLE (for sufficiently large $n$) is given by
\begin{equation*}
    e(\tilde{\pi})= \frac{I(\pi )^{-1}}{\var(\tilde{\pi})} = \frac{(1-(\pi-\pi_0))(1-\pi_0)}{1 - \pi} \leq 1,
\end{equation*}
since $\pi \geq \pi_0$. 

The efficiency loss of $\bar{\pi}$ relative to $\tilde{\pi}$ can also be expressed in terms of the increase in sample size needed when using $\bar{\pi}$ rather than $\tilde{\pi}$. Let $n^*$ denote the sample size that is needed to obtain a variance for the survey MLE $\bar{\pi}$ that is equal to the one of MME $\tilde\pi$ using a sample size of $n$. We obtain
\begin{equation*}
    \frac{n^*}{n} = \frac{1-\pi_0}{1-\frac{\pi_0}{\pi}},
\end{equation*}
which, for small $\pi_0$, is approximately equal to 
$\frac{1}{1-\pi_0/\pi}$. If, for instance, $\pi=2\pi_0$ then $\frac{n^*}{n} \approx 2$. The added value in using the additional information provided in $R_{11}$, therefore, is equivalent to using the survey MLE with a sample with twice the size.

\subsection{Confidence bounds}
\label{sec:CB}

Although the MME $\tilde{\pi}$ has a (typically small) efficiency loss relative to the conditional MLE, it has the advantage of having a known distribution through $R_{01}\sim\mathcal{B}(n,\tau_{01}(\pi))$. This allows one to construct (exact, but possibly conservative) confidence intervals even in finite samples without appealing to the estimator's asymptotic normal distribution, using the (fiducial) approach put forward in \cite{ClPe:34} (see also e.g. \citealp{fisher35,brown2001}). 

A Clopper-Pearson (CP) $(1-\gamma)$ confidence interval based on the survey MLE, i.e., based on $R_{\ast 1}$, is given by 
\begin{equation*}
    \frac{I_{\frac{\gamma}{2}}(R_{\ast 1})-\alpha}{\Delta} <\pi<\frac{I_{1-\frac{\gamma}{2}}(R_{\ast 1})-\alpha}{\Delta},
\end{equation*}
where, generally, 
\begin{eqnarray*}
    I_{\frac{\gamma}{2}}(r)= B\left(\frac{\gamma}{2}; r, n - r  + 1\right), \;\;\;\;\;
    I_{1-\frac{\gamma}{2}}(r)= B\left(1 - \frac{\gamma}{2}; r  + 1, n - r \right),
\end{eqnarray*}
and where $B(p; v, w)$, $0\leq p\leq 1$, is the cumulative distribution function of a beta distribution with shape parameters $v$ and $w$. 

A CP $(1-\gamma)$ confidence interval can be constructed based on the moment estimator \eqref{eqn:MME-ME}, i.e., based on the information provided by $R_{01}$. Given that $\mathbb{E}[R_{01}]=\pi\Delta(1-\alpha_0)-(\pi_0-\alpha_0)(1-\beta)+\alpha(1-\alpha_0)$ (see \eqref{eqn:tau-pi}), a $(1-\gamma)$ confidence interval for $\pi$, is given by
\begin{equation*}
    \frac{I_{\frac{\gamma}{2}}(R_{01})+(\pi_0-\alpha_0)(1-\beta)-\alpha(1-\alpha_0)}{\Delta(1-\alpha_0)}<\pi<\frac{I_{1-\frac{\gamma}{2}}(R_{01})+(\pi_0-\alpha_0)(1-\beta)-\alpha(1-\alpha_0)}{\Delta(1-\alpha_0)}. 
\end{equation*}

Using the conditional and marginal MLEs we can also provide confidence intervals based on their asymptotic normal distribution. All these confidence intervals are compared in our COVID-19 inspired simulation study in Section \ref{sec:simu} and in our case study using actual COVID-19 data from an Austrian survey sample in Section \ref{sec:case-study}.

\section{Simulation study}
\label{sec:simu}

In this section, we present the efficiencies, coverage and confidence interval lengths of the different methods, in finite samples. This section is parameterized in such a way that it is loosely compatible with the case of COVID-19 prevalence estimation using PCR tests. In particular, The FP and FN rates have been chosen so that they correspond to sensitivity and specificity commonly encountered in COVID-19 medical  tests, as for example reported by the Center for Health Security of the John Hopkins University \citep{KoWeGr:20}, see also \citep{surkova2020false}. Throughout we choose $\alpha_0=0$, the FP rate of the official procedure. We do so because, as pointed out in Remark \ref{rem:alpha0}, $\alpha_0$ is the product of two probabilities, here the probability of a COVID-19 negative person being selected to be tested in the official procedure and the FP rate of the PCR test employed in the official procedure. Given the relative low official prevalence of COVID-19, at least at the moment of writing this article, this product must be fairly close to zero. If, for instance, 1\% of the member of a population have been found positive through the official procedure and if the FP rate of the PCR test is another 1\%, we get an $\alpha_0=(0.01)^2=0.01\%$. 

We consider three settings. Setting I is without misclassification error, i.e.  with $\alpha_0=\alpha=\beta=0$. Setting II has only a FN rate, i.e. $\alpha_0=\alpha=0$, $\beta=2\%$. Setting III, finally, has both types of misclassification errors, i.e., $\alpha_0=0$, $\alpha=1\%$, $\beta=2\%$. We consider a sample size of $n=2,000$ which leads to the same conclusions (not presented here) as a somewhat smaller sample size (e.g. $n=1,500$). 

For $\pi$, we consider three rather different values, i.e. 5\%, 20\% and 75\% in order to cover a wide range of possible prevalence rates. For $\pi_0$, we consider, for each value of $\pi$, 30 equally spaced values between $1.025\min(\alpha_0,\pi)$ and $0.975\pi$, so that, conditionally on the information brought in by $Z_i$, one can appreciate the efficiency and accuracy gain of the approach based on the conditional model. As estimators, we consider the survey MLE $\bar{\pi}$ in \eqref{eqn:naive-ME}, the conditional MLE $\hat{\pi}$ in \eqref{eqn:MLE-ME}, the MME $\tilde{\pi}$ in \eqref{eqn:MME-ME} as well as the marginal MLE $\check{\pi}$ in \eqref{eqn:MMLE-ME} for the plausible cases when the information on $R_{10}$ and $R_{00}$ in \eqref{eqn:R-R4} is not available.

\begin{figure}
    \centering
    \includegraphics[width=0.9\textwidth]{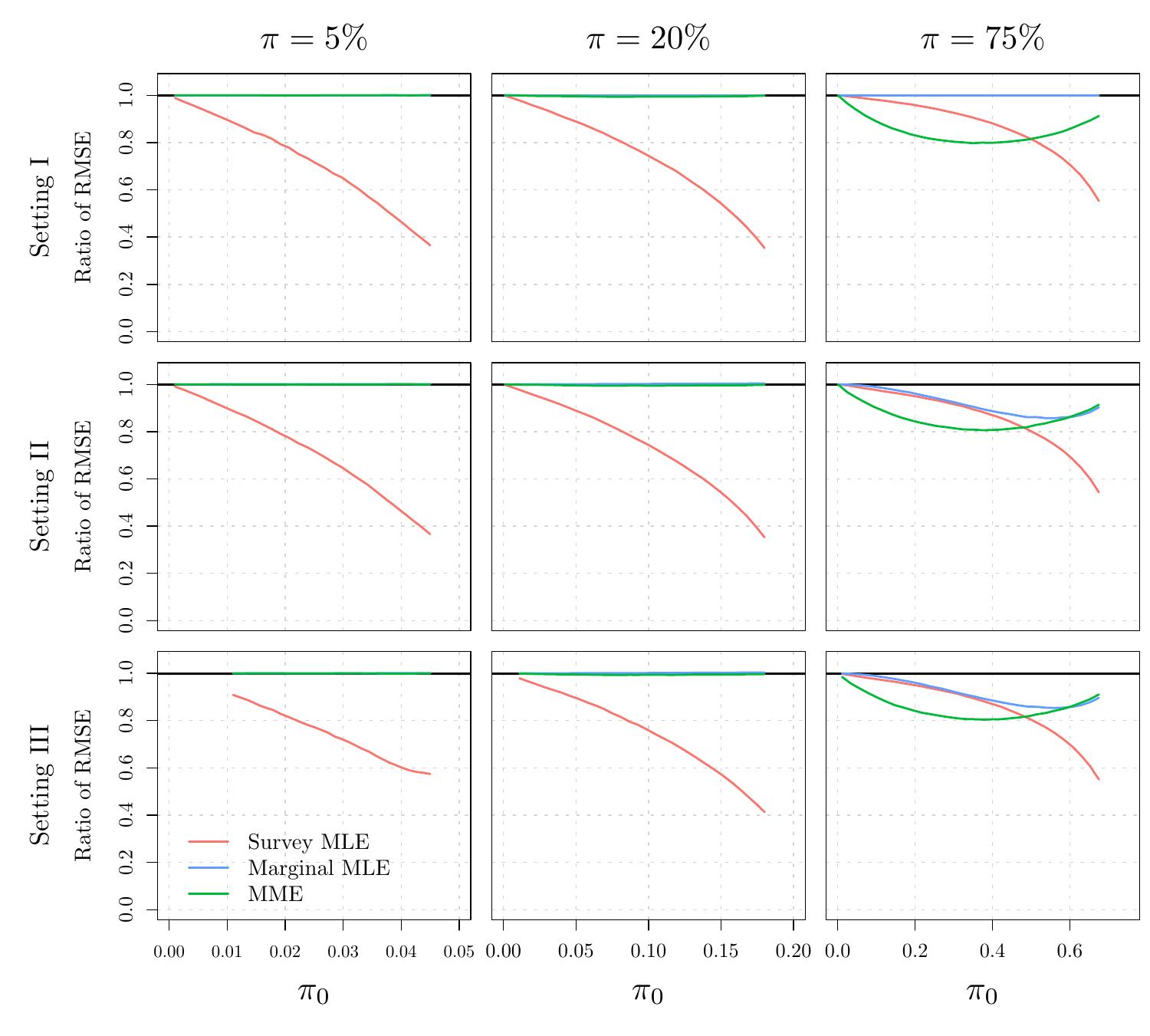}
    \caption{Relative efficiencies, as measured by the relative empirical RMSE, for the MME $\tilde{\pi}$ (green lines), the survey MLE $\bar{\pi}$ (red lines) and the marginal MLE (blue lines) relative to the conditional MLE $\hat{\pi}$. First raw with no misclassification error, middle row with FN positive rates ($\alpha_0 = \alpha = 0$, $\beta = 2\%$), bottom row with both types of misclassification errors ($\alpha_0=0$, $\alpha=1\%$, $\beta = 2\%$). The sample size is $n=2,000$ and the number of Monte Carlo simulations is $50,000$.}
    \label{fig:eff-comp}
\end{figure}

Figure \ref{fig:eff-comp} presents the relative efficiencies, as measured by the relative empirical RMSE, for the MME $\tilde{\pi}$, the survey MLE $\bar{\pi}$, and the marginal MLE $\check{\pi}$ relative to the conditional MLE $\hat{\pi}$. The main messages are the following. First, there is a substantial efficiency loss for the survey MLE $\bar{\pi}$ that increases drastically as $\pi_0$ approaches $\pi$, with or without misclassification errors. This is in line with the fact that the information brought in by considering $Z_i$ \eqref{eqn:bernoulli-rv}, is more important as $\pi_0$ is near $\pi$, and ignoring it, lowers the efficiency. Second, for the marginal MLE, the efficiency loss is negligible throughout the different settings, so there is little gain in considering $R_{10}$ and $R_{00}$ in \eqref{eqn:R-R4}, especially when this information is difficult/costly to obtain.  Third, for the MME, the efficiency loss is negligible for $\pi=5\%$ and $\pi=20\%$ when $\pi_0$ is not too near to $\pi$, while the efficiency loss is rather important for small values of $\pi_0$ (relative to $\pi$), compared to the one of the survey MLE when $\pi=75\%$. 

\begin{figure}
    \centering
    \includegraphics[width=0.9\textwidth]{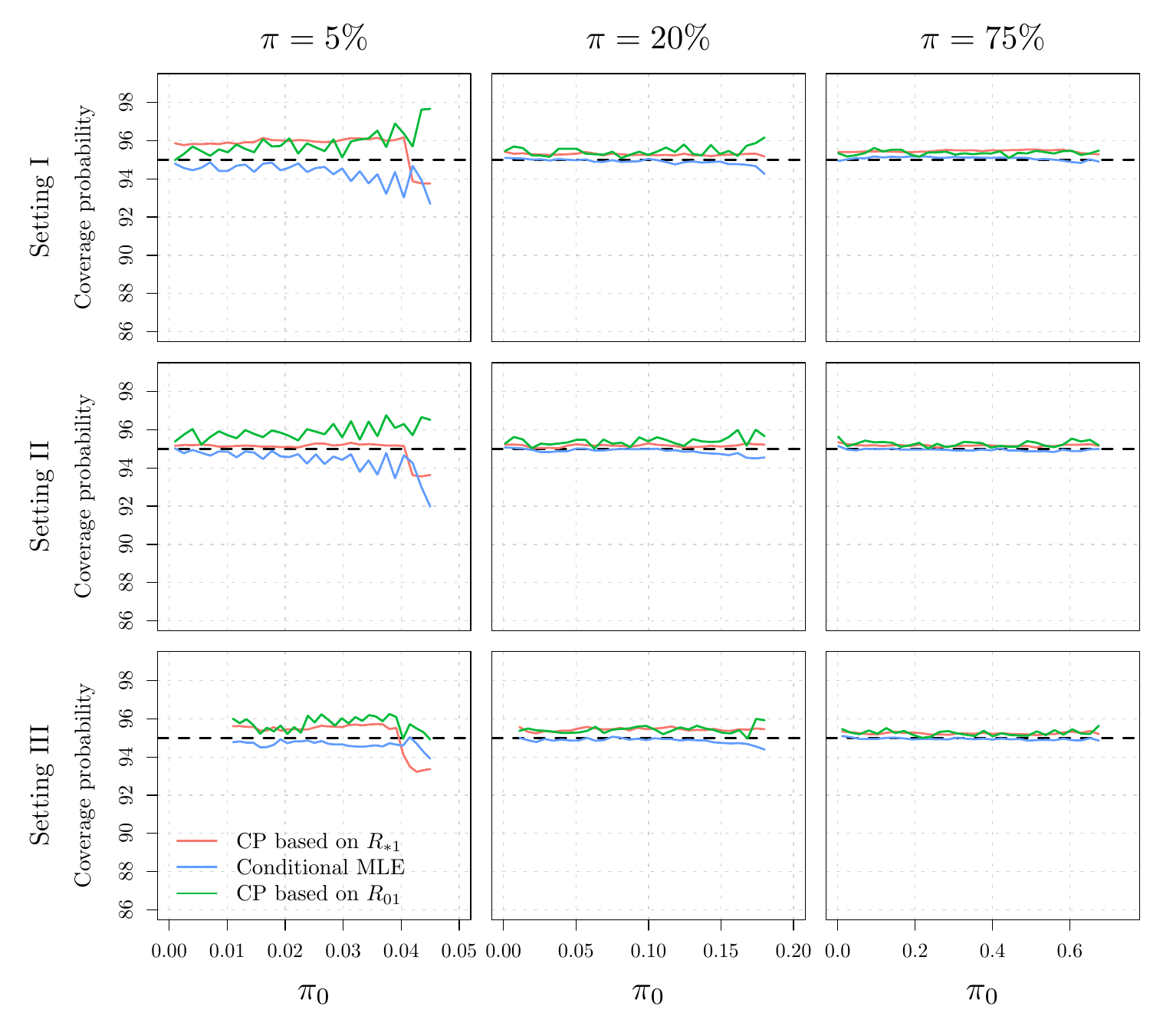}
    \caption{Empirical coverage (at the 95\% level) for the CP method based on $R_{\ast 1}$ in \eqref{eqn:R-R4}, the CP method based on $R_{01}$ in \eqref{eqn:R-R4} and the asymptotic method based on the conditional MLE $\hat\pi$. Top panels: $\alpha_0=\alpha=\beta=0$. Middle panels: $\alpha_0 = \alpha = 0$, $\beta = 2\%$. Bottom panels: $\alpha_0=0$, $\alpha=1\%$, $\beta = 2\%$. The sample size is $2,000$ and the number of Monte Carlo simulations is $50,000$.}
    \label{fig:coverage}
\end{figure}

Figure \ref{fig:coverage} presents the coverage (at the 95\% level), computed using simulations, for the CP method based on $R_{\ast 1}$ in \eqref{eqn:R-R4}, which is associated to the survey MLE $\bar\pi$, the CP method based on $R_{01}$ in \eqref{eqn:R-R4}, which is associated to the MME $\tilde\pi$, and the asymptotic method based on the conditional MLE $\hat\pi$. The coverage for the asymptotic method based on the marginal MLE $\check\pi$ are not presented as they are the same as the ones for the asymptotic method based on the conditional MLE. Overall, as expected, the CP method provides slightly conservative coverage across settings, while the asymptotic method based on the survey MLE is slightly liberal, especially for $\pi=5\%$. Moreover, for both the CP method based on $R_{01}$ and the asymptotic method based on the conditional MLE, for $\pi=5\%$ and $\pi=20\%$, the coverage worsens (even if they remain quite accurate) as $\pi_0$ approaches $\pi$. For the asymptotic method, this can be explained by the fact that confidence intervals might have bounds falling outside the domain of $\pi$ (e.g. below $\pi_0$), especially when $\pi$ is near $\pi_0$ and in settings such as Setting II.

\begin{figure}
    \centering
    \includegraphics[width=0.9\textwidth]{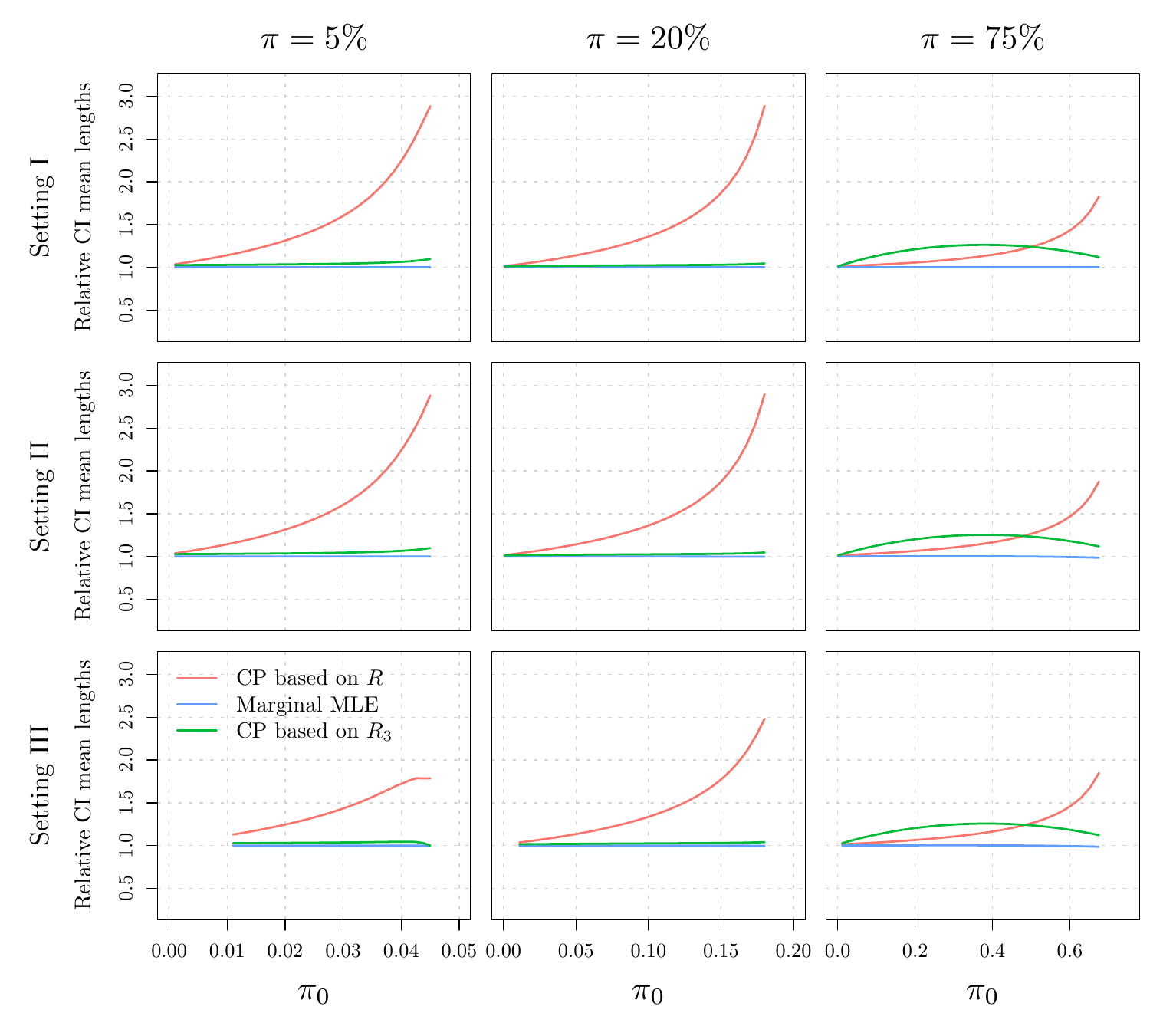}
    \caption{Relative empirical confidence interval (at the 95\% level) mean lengths for the CP method based on $R_{\ast 1}$ in \eqref{eqn:R-R4} and the CP method based on $R_{01}$ in \eqref{eqn:R-R4}, relative to the empirical confidence interval (at the 95\% level) mean lengths for the asymptotic method based on the conditional MLE $\hat\pi$.  Top panels: $\alpha_0=\alpha=\beta=0$. Middle panels: $\alpha_0 = \alpha = 0$, $\beta = 2\%$. Bottom panels: $\alpha_0=0$, $\alpha=1\%$, $\beta = 2\%$. The sample size is $2,000$ and the number of Monte Carlo simulations is $50,000$.}
    \label{fig:CIlenght}
\end{figure}

Given that the coverage is reasonable across methods, it is worth comparing the confidence interval lengths. Figure \ref{fig:CIlenght} presents the relative confidence interval (at the 95\% level) lengths, computed using simulations, for the CP method based on $R_{\ast 1}$ in \eqref{eqn:R-R4} (associated to the survey MLE $\bar\pi$) and the CP method based on $R_{01}$ in \eqref{eqn:R-R4} (associated to the MME $\tilde\pi$), relative to the confidence interval (at the 95\% level) lengths for the asymptotic method based on the conditional MLE $\hat\pi$. One can observe, as expected, that the (mean) confidence interval lengths can be a lot larger when ignoring the information provided by $Z_i$ in \eqref{eqn:bernoulli-rv}, especially as the information increases, i.e. as $\pi_0$ approaches $\pi$. An interesting feature appears, however, for a small population proportion ($\pi=5\%$) when $\pi_0$ approaches $\pi$, in that the mean confidence interval length for the CP based on $R_{01}$ (associated to the MME) is smaller than the one of the asymptotic method based on the conditional MLE. However, for a large population proportion ($\pi=75\%$), the mean confidence interval length for the CP based on $R_{\ast 1}$ are relatively smaller than the ones based on $R_{01}$, while remaining larger than the mean confidence interval length for the asymptotic method based on the conditional MLE. This is especially the case for small values of $\pi_0$ relative to $\pi$, and is in line with the study of the efficiencies provided in Figure \ref{fig:eff-comp}.

\section{Case study: Application to Austrian COVID-19 survey}
\label{sec:case-study}

We use the methodology developed in this paper for the case of the COVID-19 prevalence estimation using the results of a survey done in  November 2020 by \cite{StatAU:20}. We also compare the different approaches, in order to illustrate, in practice, the impact of choosing one method rather than another one. In November 2020, a survey sample of $n=2287$ was collected to test for COVID-19 using PCR-tests. Seventy-one participants ($R_{\ast 1}=71$) were tested positive, and among these ones, thirty-two ($R_{11}=32$) had declared to have been tested positive with the official procedure, during the same month. In November, there were $93914$  declared cases among the official (approximately) $7166167$ inhabitants in Austria (above 16 years old), so that $\pi_0 \approx 1.3105\%$. The sensitivity ($1-\alpha$) and the specificity ($1-\beta$) are not known with precision, so that we present estimates of the prevalence without misclassification error as well as for values for the FP and FN rates, that are plausible given the data and according to the sensitivity and specificity reported in \cite{KoWeGr:20} or \cite{surkova2020false}.

Table \ref{tab:preval-Austria} provides various estimates of $\pi$, the COVID-19 prevalence in Austria in November 2020, for the case of no misclassification error and for the case of misclassification errors with $\alpha  = 1\%$, $\beta = 10\%$, and $\alpha_0 = 0$. Recall Remark~\ref{rem:alpha0} for the choice of $\alpha_0=0$. We also chose a small $\alpha$ (FP rate for the medical test in the survey sample), because we only observe 71 positive cases out 2287 participants. If $\alpha$ were larger, say $\alpha=5\%$, we would also expect a larger number of (misclassified) positive cases, i.e. $114$ positive cases just because of false positives. 

From the first three lines of Table \ref{tab:preval-Austria}, one can derive a series of insights. First, we note that without misclassification errors, the estimates are very similar across methods. Second, as expected, the confidence intervals for the SMLE are wider than the ones associated to the conditional MLE (CMLE) or the MME. These two statements are true for both the case of no misclassification errors and the case of some misclassification errors.  Third, in the case of misclassification errors, the estimates differ more substantially between the sample MLE and the conditional MLE or MME, with a difference of $10\%$ in the estimate. 

\begin{figure}
    \centering
    \includegraphics[width=0.75\textwidth]{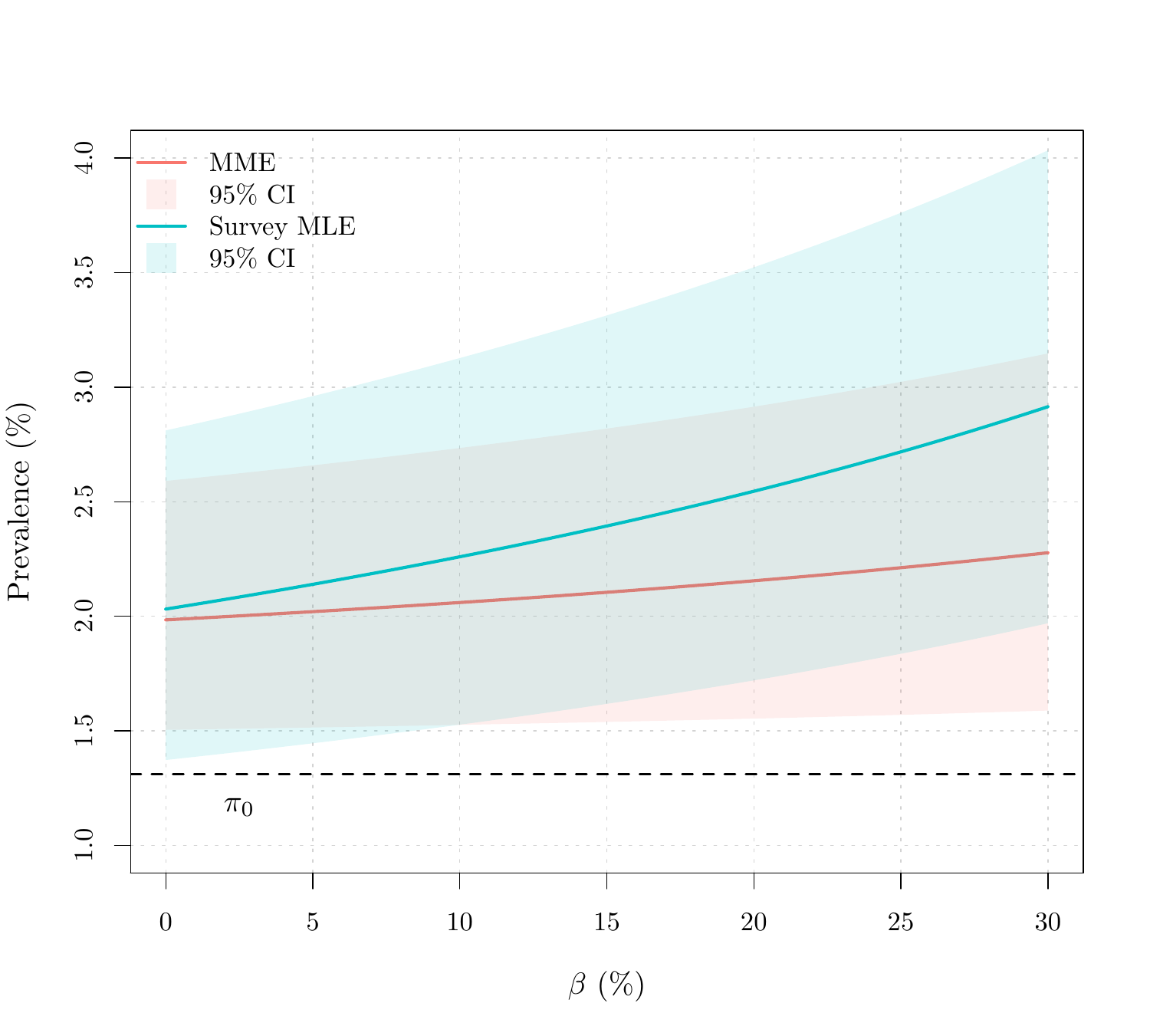}
    \caption{Sensitivity analysis for the prevalence estimation using the moment estimator and the survey MLE, according the the FP rate $\beta$. The confidence bounds are computed using the CP-method for both estimators. $\alpha_0=0$, $\alpha=1\%$, $\pi_0=1.3105\%$, $n=2287$, $R_{\ast 1}=71$ and $R_{11}=32$.}
    \label{fig:beta-variation}
\end{figure}

Since the FP rate $\alpha$ has a limited number of possible values, given the data, we present in Figure \ref{fig:beta-variation} a sensitivity analysis of the prevalence estimation by the survey MLE and the MME, when the FN $\beta$ varies from $0\%$ to $30\%$. What is striking is that the sample MLE is much more influenced by the value of the FP $\beta$ compared to the MME, which shows a far better stability. To understand this feature, from \eqref{eqn:MME-ME}, we get, for the MME, under the sensitivity analysis conditions, $\tilde{\pi} = \frac{1}{\Delta} \left(\frac{R_{01}}{n} + (1-\beta)\pi_0 \right)$. With increasing values for $\beta$, $\Delta=1-(\alpha+\beta)$ decreases, but at the same time, the quantity $(1-\beta)\pi_0$ also decreases. On the other hand, with the survey MLE given in \eqref{eqn:naive-ME}, an increase in the FP $\beta$ directly induces an increased value for the estimator.

Finally, in order to illustrate the accuracy gain of using a conditional MLE or MME, in Table \ref{tab:preval-Austria}, last two lines, we provide the prevalence estimate using the sample MLE with associated CI with 1.5 and 2 times as many sample data. In other words, the (hypothetical) data are built up by choosing $n^*=\ceil{kn}$ and $R_{*1}^*=\ceil{kR_{*1}}$, with $k=1.5,2$. The aim of this exercise it to see if with more data, the sample MLE can provide an estimator that is as accurate as the conditional MLE or MME. One can see that, roughly, one would need twice as much survey sample data, in order to achieve the same level of accuracy provided by the MME or the conditional MLE. This is in line with the theoretical results provided in Section \ref{sec:efficiency}.

%Moreover, for larger values of $\beta$, which might not be realistic, the MME estimate can lye outside the confidence interval of the survey MLE.
%Third, we can see that, in the case of no misclassification errors, the confidence interval based on the asymptotic distribution of the UMLE as well as that of the CMLE have a lower bound beyond the (known) proportion $\pi_0=0.132\%$ of positive cases found through the official procedure. has a lower bound that is below $\pi_0=0.132\%$. Fifth, this is however not the case (with the Clopper-Pearson confidence interval) when considering the MME, which, by definition, has to have a lower bound that is at least as great as $\pi_0=0.132\%$. This is, again, for the case of no classification errors. For the case of misclassification errors, the Clopper-Pearson confidence interval can also have a lower bound below the known proportion of $\pi_0=0.132\%$. 

\begin{table}[]
\footnotesize
    \centering
    \begin{tabular}{lccc@{}ccc@{}}
    \toprule
     & \multicolumn{3}{c}{$\alpha_0=\alpha=\beta=0$} & \multicolumn{3}{c}{$\alpha_0=0,\alpha=1\%,\beta=10\%$}\\
     \cmidrule(r){2-4}  \cmidrule(r){5-7} & Estimates (\%) & 95\% CI (\%) & Illustration & Estimates (\%) & 95\% CI (\%)  & Illustration \\ \midrule
     %SMLE-as & $3.011$ & $(2.321 - 3.701)$ & {\raisebox{-.3\height}{\includegraphics[width=1.5cm]{fig22.pdf}}} & $2.260$ & $(1.485 - 3.035)$ & {\raisebox{-.3\height}{\includegraphics[width=1.5cm]{fig21.pdf}}}  \\
      CMLE-as & $2.965$ & $(2.450 - 3.480)$ & {\raisebox{-.3\height}{\includegraphics[width=2.3cm, height = 0.58cm]{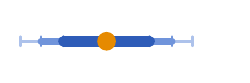}}} & $2.062$ & $(1.484 - 2.641)$ & {\raisebox{-.3\height}{\includegraphics[width=2.3cm, height = 0.58cm]{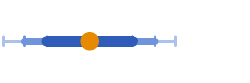}}}\\
      MME-CP & $2.965$ & $(2.489 - 3.565)$ & {\raisebox{-.3\height}{\includegraphics[width=2.3cm, height = 0.58cm]{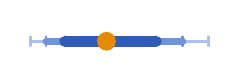}}}& $2.060$ & $(1.526 - 2.734)$ & {\raisebox{-.3\height}{\includegraphics[width=2.3cm, height = 0.58cm]{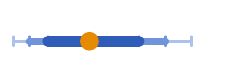}}}\\
      SMLE-CP & $3.011$ & $(2.359 - 3.783)$ & {\raisebox{-.3\height}{\includegraphics[width=2.3cm, height = 0.58cm]{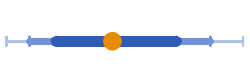}}}& $2.260$ & $(1.527 - 3.127)$ & {\raisebox{-.3\height}{\includegraphics[width=2.3cm, height = 0.58cm]{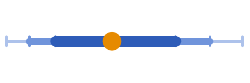}}}\\
      SMLE-CP$^*$ & $3.011$ & $(2.486 - 3.644)$ & {\raisebox{-.3\height}{\includegraphics[width=2.3cm, height = 0.58cm]{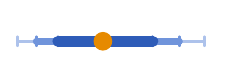}}}& $2.260$ & $(1.669 - 2.971)$ & {\raisebox{-.3\height}{\includegraphics[width=2.3cm, height = 0.58cm]{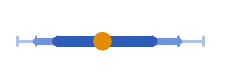}}}\\
      SMLE-CP$^{**}$ & $3.011$ & $(2.542 - 3.539)$ & {\raisebox{-.3\height}{\includegraphics[width=2.3cm, height = 0.58cm]{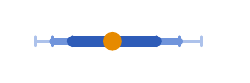}}}& $2.260$ & $(1.733 - 2.853)$ & {\raisebox{-.3\height}{\includegraphics[width=2.3cm, height = 0.58cm]{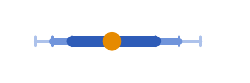}}} \\
      %\midrule 
      %& \multicolumn{5}{l}{$^*$: $n^*=1.5n$, $R_{*1}^*=1.5R_{*1}$.} \\
      %& \multicolumn{5}{l}{$^{**}$: $n^*=2n$, $R_{*1}^*=2R_{*1}$.} \\
    \bottomrule \vspace{0.01cm}
    \end{tabular}
    \caption{Prevalence estimation for the Austrian data (November 2020) with associated $95$\% confidence intervals, using the conditional MLE (CMLE) with asymptotic confidence intervals, the moment estimator (MME) with Clopper-Pearson intervals and the survey MLE (SMLE) with asymptotic and Clopper-Pearson confidence intervals. For the later, two additional estimation are provided with $n^*=\ceil{kn}$ and $R_{*1}^*=\ceil{kR_{*1}}$, with $k=1.5$ (SMLE-CP$^*$) and $k=2$ (SMLE-CP$^{**}$). The original data are $\pi_0 \approx 1.3105\%$, $n=2287$, $R_{\ast 1}=71$, $R_{11}=32$. The CI are illustrated as horizontal bars with lengths associated  to respectively $80\%$, $95\%$ and $99\%$ confidence levels, with a dot representing the estimate.   The first three columns are under the assumption of no misclassification errors. The second three columns assume $\alpha  = 1\%$, $\beta = 10\%$, and $\alpha_0 = 0$.} 
    \label{tab:preval-Austria}
\end{table}

\section{Conclusions}

While we have cast this paper in the language of disease prevalence estimation, the method we propose has a more general range of applications. We actually propose a method to estimate the proportion of some characteristic in a population using information both from a random sample and from an incomplete census or census with participation bias. In other words, we are interested in the prevalence (or proportion) of population members having characteristic A, conditional on another characteristic B, such that having characteristic B implies having characteristic A, but not necessarily vice versa. We study this problem with and without the possibility of misclassification errors for A as well as for B.

The approach that we propose for such settings is that when a random survey sample is drawn to not only record for each participant whether they have characteristic A or not, but also whether they have characteristic B or not. The key idea, to improve the accuracy of the estimate of the prevalence of characteristic A in the population, is to base the estimate appropriately on the number of participants in the sample that have characteristic A and not B. We propose MLE as well as MME derived from this idea. 

We show that our approach provides estimates that are substantially more accurate than the simple sample proportion (of participants with characteristic A), the maximum likelihood estimate that ignores the information available for characteristic B. As an important consequence, our approach can provide a given level of desired accuracy, with a substantially smaller sample size. This is useful when data collection is costly or, as for our COVID-19 example, medical tests (or lab spaces to evaluate test) are in limited supply.

It would be straightforward to adapt the estimators to the case of weighted sampling, with non random weights, as well as to
include explanatory variables in our model in the same vein as in generalized linear models by postulating a relationship of the proportion parameter of interest and an array of additional observable characteristics. 

Finally, there is some similarity of our approach and that of capture-recapture models \citep[see e.g.][and the references therein]{ChaoCapture:2001} used to estimate the size of a population. In capture-recapture models several samples are drawn randomly from a population with unknown size. Estimates of the size then, as in our approach, rely on the possibility of participants in a first sample showing up again in a second sample. To see the difference between the two approaches, we can place our framework in the language of capture-recapture models as follows. In our case, the first capture is taken as an incomplete census or a census with participation bias from a population of known size, and not a random sample from a population of unknown size.

%The estimation approach proposed in this paper can be used to measure prevalence in the presence of misclassification error when the true prevalence is small, using the information provided by misclassification taken in a sub-population. It has the advantage of being more accurate. One can also extend the framework to the case when the distributions of the sensitivity and specificity are known, by introducing them (instead of cutoff point) in the likelihood function \citep[see e.g.][]{Bouman2020}, but this extension is left for further research.

\section{Software}
\label{sec:cape}

All computations presented in this paper were done using the ContionAl Prevalence Estimation, or \textrm{cape R} package that can be downloaded from \url{https://github.com/stephaneguerrier/cape}. Installation instructions as well as a user guide (vignette) of the package are provided in \url{https://stephaneguerrier.github.io/cape/}. All simulation results (as well as additional ones), can be reproduced and the simulation script is available on GitHub. 

\section*{Acknowledgments}
St\'ephane Guerrier is partially supported by Swiss National Science Foundation grant \#176843 and Innosuisse-Boomerang Grant 37308.1 IP-ENG. Maria-Pia Victoria-Feser is partially supported by a Swiss National Science Foundation grant \#182684. We are grateful to Michael Greinecker, Helmut Kuzmics, Hans Manner, Michael Richter, Michael Scholz and Dominique-Laurent Couturier for helpful comments and suggestions.

\newpage
\bibliographystyle{chicago}  
\bibliography{refs}

% ------------------------------------
% --- Appendix
% ------------------------------------
\newpage
\appendix
\centerline{\Large\sc Supplementary Material}

\section{Success probabilities} 
\label{app:succprob}

The success probabilities $\tau_j(\pi)$ for $R_j$, $j=1,2,3,4$, in \eqref{eqn:R-R4}, can be deduced from the following table. There are two fundamental cases $X_i=0$ and $X_i=1$, and conditionally on each one of these cases, errors are independently and identically distributed. 

\[\begin{array}{cc|cc}
X_i & prob & \mathbb{P}(Y_i=1) & \mathbb{P}(Z_i=1) \\ \hline
1 & \pi & 1-\beta & 1-\beta_0 \\
0 & 1-\pi & \alpha & \alpha_0, 
\end{array}
\]
where $\beta_0 = 1- \frac{\pi_0-\alpha_0(1-\pi)}{\pi}$.

We, thus, have
\begin{eqnarray*}
\tau_{11} & = & \mathbb{P}(X_i=1) \mathbb{P}(Y_i=1 \vert X_i=1) \mathbb{P}(Z_i=1 \vert X_i=1) + \mathbb{P}(X_i=0) \mathbb{P}(Y_i=1 \vert X_i=0) \mathbb{P}(Z_i=1 \vert X_i=0) \\
& = & \pi (1-\beta) (1-\beta_0) + (1-\pi) \alpha \alpha_0.
\end{eqnarray*}
Plugging in $\beta_0 = 1- \frac{\pi_0-\alpha_0(1-\pi)}{\pi}$ and using $\Delta \vcentcolon=1-(\alpha+\beta)$ we obtain 
\[\tau_{11}(\pi) = \pi\Delta\alpha_0+(\pi_0-\alpha_0)(1-\beta)+\alpha\alpha_0 .\] 

The remaining probabilities $\tau_{10},\tau_{01}$ and $\tau_{00}$ can be similarly obtained. 

%\ckcomment{Do we need this? If so, do we want more steps? Do we want to do this for all $\tau$'s?}
%\sgcomment{I think this can be useful but I don't think we need more steps.}

\section{Proof of Proposition \ref{prop:mle}}
\label{app:proof-prop:mle}

\begin{proof}
The identifiability of the model is straightforward from \eqref{eqn:tau-pi} and by the extreme value theorem we have $\mathbb{E}[|\ln p(\mathbf{R} | \pi) |] < \infty$, where $p(\mathbf{R} | \pi)$ denotes the probability mass function of a multinomial distribution with event probabilities $\tau_i, \, i = 1, 2, 3, 4$ as defined in \eqref{eqn:tau-pi}. Therefore, by applying the information inequality (see e.g. Lemma 2.2 of \citealp{newey1994large}), we can verify the identification of $\hat{\pi}$. By combining the compactness of $\Pi$, the (uniform) law of large numbers and/or Theorem 2.1. of \cite{newey1994large}, $\hat{\pi}$ is a consistent estimator for $p_0$. Then, if $p_0 \in (\underline{\pi}, 1)$, standard techniques can be used to show that
\begin{equation*}
    \sqrt{n} \left(\hat{\pi} - p_0\right) \xrightarrow[n \to \infty]{\mathcal{D}} \mathcal{N} \left(0, \frac{1}{I(p_0)}\right),
\end{equation*}
where
\begin{equation*}
\begin{aligned}
     I(\pi) &= \Delta^2 \Bigg( \frac{\alpha_0^2 \tau_{11}(\pi)}{\left[\alpha \alpha_0 + (1 - \beta) (\pi_0 - \alpha_0) + \pi\alpha_0 \Delta\right]^2} +
     \frac{\alpha_0^2 \tau_{10}(\pi)}{\left[\beta (\pi_0 - \alpha_0) + \alpha_0 \left(1 - \pi \Delta - \alpha\right)\right]^2}\\
     &+
     \frac{\left(1 - \alpha_0\right)^2 \tau_{01}(\pi)}{\left[(1 - \beta) (\pi_0 - \alpha_0) - \pi \left(1 - \alpha_0\right) \Delta - \alpha (1 - \alpha_0)\right]^2} + 
     \frac{\left(1 - \alpha_0\right)^2 \tau_{00}(\pi)}{\left[(1 - \alpha_0) \left(1 - \pi \Delta - \alpha \right) - \beta (\pi_0 - \alpha_0)\right]^2} \Bigg).
\end{aligned}
\end{equation*}
Finally, we verify that Assumption \ref{ass:alpha-beta} guarantee that $I(\pi)$ exists and is finite. Indeed, none of the equations:
\begin{equation*}
    \begin{aligned}
    \alpha \alpha_0 + (1 - \beta) (\pi_0 - \alpha_0) + \pi\alpha_0 \Delta &= 0\\
    \beta (\pi_0 - \alpha_0) + \alpha_0 \left(1 - \pi \Delta - \alpha\right)&= 0\\
    (1 - \beta) (\pi_0 - \alpha_0) - \pi \left(1 - \alpha_0\right) \Delta - \alpha (1 - \alpha_0)&= 0\\
    (1 - \alpha_0) \left(1 - \pi \Delta - \alpha \right) - \beta (\pi_0 - \alpha_0)&= 0,
    \end{aligned}
\end{equation*}
have a solution in $(\underline{\pi}, 1)$, which concludes the proof.
\end{proof}

\section{Alternative GMM estimators for the conditional model}
\label{app:gmm}

A possibly more efficient and closed form estimator can be obtained by choosing a weighted sum of the $\tilde{\pi}^{(l)}$, with weights summing to one to obtain an unbiased estimator with a smaller variance. Indeed, let for example $\omega_{ll}=\gamma_l, l=1,2,3$ and $0$ otherwise, such that 
\begin{equation}
    \ddot{\pi}(\bgam) = \sum_{l=1}^3 \gamma_l \tilde{\pi}^{(l)}
    \label{eqn:GMM-gamma}
\end{equation}
and $\bgam=[\gamma_l]_{l=1,2,3}$, with $\sum_{l=1}^3 \gamma_l = 1$, we can choose $\bgam$ such that
\begin{equation*}
    \min_{\bgam } \var(\ddot{\pi}(\bgam)).
\end{equation*}
The fourth term $l=4$ is omitted as it does not provide additional information, since we have that $\sum_{i=0}^1 \sum_{j=0}^1 R_{ij}=n$. As is shown below, we have that
\begin{eqnarray}
    \gamma_1&=&\frac{\lambda \alpha_0}{2}\left( \frac{1-\alpha_0}{\tau_{00}} + \frac{\alpha_0}{\tau_{11}} \right) \nonumber \\
    \gamma_2&=&\frac{\lambda \alpha_0}{2}\left( \frac{\alpha_0}{\tau_{10}} - \frac{1-\alpha_0}{\tau_{00}} \right) \label{eqn:gamma_opt} \\
    \gamma_3&=&\frac{\lambda (1-\alpha_0)}{2}\left( \frac{1-\alpha_0}{\tau_{00}} + \frac{1-\alpha_0}{\tau_{01}} \right). \nonumber
\end{eqnarray}
One can see that the weight $\gamma_3$ is the most important, as $\alpha_0$ is usually very small, see Remark \ref{rem:alpha0}. Unfortunately, the weights $\bgam$ depend on $\pi$, so that one needs to plug in a value. This could be chosen as being the one provided by $\tilde{\pi}$ in \eqref{eqn:MME-ME}, which is a consistent estimator of $\pi$. Nevertheless, the finite sample distribution of $\tilde{\pi}(\bgam)$ in \eqref{eqn:GMM-gamma} is unknown, so that one would need to resort to asymptotic theory, and this would not bring any advantage, in terms of inference, compared to the MLE.

To obtain \eqref{eqn:gamma_opt}, we first develop \eqref{eqn:pi-l} using \eqref{eqn:tau-pi} to obtain
\begin{eqnarray*}
\tilde{\pi}_1 &=& \frac{1}{\Delta \alpha_0} \left(\frac{R_{11}}{n} - (\pi_0 - \alpha_0)(1- \beta) - \alpha \alpha_0 \right) \\
\tilde{\pi}_2 &=& \frac{1}{\Delta \alpha_0} \left((\pi_0 - \alpha_0) \beta + (1-\alpha) \alpha_0 - \frac{R_{10}}{n}\right) \\
\tilde{\pi}_3 &=& \frac{1}{\Delta(1-\alpha_0)} \left(\frac{R_{01}}{n} + \pi_0 - \beta\pi_0 - \alpha_0 \Delta - \alpha\right).
\end{eqnarray*}
Letting $\tau_j\vcentcolon =\tau_j(\pi),j=1,\ldots,4$, the variance of the GMM $\ddot\pi$ in \eqref{eqn:GMM-gamma}, using the properties of the multinomial distribution, is given by
\begin{eqnarray*}
\var(\ddot{\pi}) & = & \frac{\gamma_1^2}{n\Delta^2\alpha_0^2} \tau_{11} (1-\tau_{11}) + \frac{\gamma_2^2}{n\Delta^2\alpha_0^2} \tau_{10} (1-\tau_{10}) + \frac{\gamma_3^2}{n\Delta^2(1-\alpha_0)^2} \tau_{01} (1-\tau_{01}) \\ &  & + 2 \frac{\gamma_1 \gamma_2}{n\Delta^2\alpha_0^2} \tau_{11} \tau_{10} - 2 \frac{\gamma_1 \gamma_3}{n\Delta^2\alpha_0(1-\alpha_0)} \tau_{11} \tau_{01} + 2 \frac{\gamma_2 \gamma_3}{n\Delta^2\alpha_0(1-\alpha_0)} \tau_{10} \tau_{01}.
\end{eqnarray*}
%
%\begin{eqnarray*}
%\var(\ddot{\pi}) & = & \frac{\gamma_1^2}{\Delta^2\alpha_0^2} \var(R_{11}/n) + \frac{\gamma_2^2}{\Delta^2\alpha_0^2} \var(R_{10}/n) + \frac{\gamma_3^2}{\Delta^2(1-\alpha_0)^2} \var(R_{11}/n) \\ &  & - 2 \frac{\gamma_1 \gamma_2}{\Delta^2\alpha_0^2} \cov(R_{11}/n,R_{10}/n) + 2 \frac{\gamma_1 \gamma_3}{\Delta^2\alpha_0(1-\alpha_0)} \cov(R_{11}/n,R_{01}/n) - 2 \frac{\gamma_2 \gamma_3}{\Delta^2\alpha_0(1-\alpha_0)} \cov(R_{10}/n,R_{01}/n),
%\end{eqnarray*}
%
Minimizing the variance subject to $\sum_j \gamma_j=1$ is then equivalent to minimizing
\begin{eqnarray*}
\mathcal{H}(\gamma) & = & \frac{\gamma_1^2}{\alpha_0^2} \tau_{11} (1-\tau_{11}) + \frac{\gamma_2^2}{\alpha_0^2} \tau_{10} (1-\tau_{10}) + \frac{\gamma_3^2}{(1-\alpha_0)^2} \tau_{01} (1-\tau_{01}) \\ &  & + 2 \frac{\gamma_1 \gamma_2}{\alpha_0^2} \tau_{11} \tau_{10} - 2 \frac{\gamma_1 \gamma_3}{\alpha_0(1-\alpha_0)} \tau_{11} \tau_{01} + 2 \frac{\gamma_2 \gamma_3}{\alpha_0(1-\alpha_0)} \tau_{10} \tau_{01} \\
& & - \lambda(\gamma_1+\gamma_2+\gamma_3-1).
\end{eqnarray*}
The first order conditions for minimality are then given by
\begin{eqnarray*}
\frac{\partial \mathcal{H}}{\partial \gamma_1} & = & \frac{2\gamma_1}{\alpha_0^2} \tau_{11} (1-\tau_{11}) + \frac{2\gamma_2}{\alpha_0^2} \tau_{11} \tau_{10} - \frac{2\gamma_3}{\alpha_0(1-\alpha_0)} \tau_{11} \tau_{01} - \lambda = 0 \\
\frac{\partial \mathcal{H}}{\partial \gamma_2} & = & \frac{2\gamma_2}{\alpha_0^2} \tau_{10} (1-\tau_{10}) + \frac{2\gamma_1}{\alpha_0^2} \tau_{11} \tau_{10} + \frac{2\gamma_3}{\alpha_0(1-\alpha_0)} \tau_{10} \tau_{01} - \lambda = 0 \\
\frac{\partial \mathcal{H}}{\partial \gamma_3} & = & \frac{2\gamma_3}{(1-\alpha_0)^2} \tau_{01} (1-\tau_{01}) - \frac{2\gamma_1}{\alpha_0(1-\alpha_0)} \tau_{11} \tau_{01} + \frac{2\gamma_2}{\alpha_0(1-\alpha_0)} \tau_{10} \tau_{01} - \lambda = 0 \\
\end{eqnarray*}
which can be simplified as
\begin{eqnarray} 
\frac{2\gamma_1}{\alpha_0} (1-\tau_{11}) + \frac{2\gamma_2}{\alpha_0} \tau_{10} - \frac{2\gamma_3}{1-\alpha_0} \tau_{01} & = & \frac{\lambda \alpha_0}{2\tau_{11}}  \label{eqn:firsteq} \\
\frac{2\gamma_2}{\alpha_0} (1-\tau_{10}) + \frac{2\gamma_1}{\alpha_0} \tau_{11} + \frac{2\gamma_3}{1-\alpha_0} \tau_{01} & = & \frac{\lambda \alpha_0}{2\tau_{10}} \label{eqn:secondeq} \\
\frac{2\gamma_3}{1-\alpha_0} (1-\tau_{01}) - \frac{2\gamma_1}{\alpha_0} \tau_{11} + \frac{2\gamma_2}{\alpha_0} \tau_{10} & = & \frac{\lambda (1-\alpha_0)}{2\tau_{01}} \label{eqn:thirdeq}
\end{eqnarray}
Using \eqref{eqn:firsteq}  in \eqref{eqn:secondeq} to simplify for $\gamma_3$ yields
\begin{equation}
    \frac{\gamma_2}{\alpha_0} + \frac{\gamma_1}{\alpha_0} = \frac{\lambda \alpha_0}{2}\left(\frac{1}{\tau_{11}} + \frac{1}{\tau_{10}} \right).
    \label{eqn:simp1}
\end{equation}
Similarly using \eqref{eqn:firsteq} in \eqref{eqn:thirdeq} leads to 
\begin{equation}
    \frac{\gamma_1}{\alpha_0}\left(1-\tau_{11}-\tau_{01}\right) + \frac{\gamma_2}{\alpha_0} \tau_{10} = \frac{\lambda}{2}\left(1-\alpha_0+\frac{(1-\tau_{01})\alpha_0}{\tau_{11}}\right).
    \label{eqn:simp2}
\end{equation}
Then, from \eqref{eqn:simp1} and \eqref{eqn:simp2}, knowing that $\sum_{i=0}^1 \sum_{j=0}^1 \tau_{ij}=1$, we obtain
\[\frac{\gamma_1}{\alpha_0}\tau_{00} = \frac{\lambda}{2}\left(1+\frac{\alpha_0}{\tau_{11}}(\tau_{00}-\tau_{11})\right),\]
which leads to $\gamma_1$ in \eqref{eqn:gamma_opt}. Using $\gamma_1$ in e.g. \eqref{eqn:simp1}, we obtain 
$\gamma_2$ in \eqref{eqn:gamma_opt}, and finally $\gamma_3$ is deduced as in \eqref{eqn:gamma_opt}.

\end{document}